\pdfoutput=1

\documentclass[aps,prb,reprint,showpacs,superscriptaddress]{revtex4-1}
\usepackage{graphicx}
\usepackage{amsmath}
\usepackage{amssymb}
\usepackage{amsfonts}
\usepackage{dcolumn}
\usepackage{dsfont}
\usepackage{latexsym}
\usepackage{rotating}
\usepackage{color}
\usepackage{latexsym}
\usepackage{bbm}
\usepackage{subfigure}
\usepackage{float}
\usepackage{epsfig}
\usepackage{psfrag}
\usepackage{natbib}
\usepackage{bm}
\usepackage{amsthm}
\usepackage{eucal}
\usepackage{url}

\usepackage{color} 

\usepackage{hyperref}
\hypersetup{
colorlinks=true,final=true,
        linkcolor=blue,
        citecolor=blue,
        filecolor=blue,
        urlcolor=blue,
}

\begin{document}

\title{Effect of oxygen vacancy on structural, electronic and magnetic properties of La-based oxide interfaces}
\author{Urmimala Dey}
\affiliation{Centre for Theoretical Studies, Indian Institute of Technology, Kharagpur, India}
\author{Swastika Chatterjee}
\affiliation{Department of Physics, Indian Institute of Technology, Kharagpur, India}
\author{A.Taraphder}
\affiliation{Department of Physics, Indian Institute of Technology, Kharagpur, India}
\affiliation{Centre for Theoretical Studies, Indian Institute of Technology, Kharagpur, India}

\date{\today}


\begin{abstract}
{

Disorder, primarily in the form of oxygen vacancies, cation stoichiometry and atomic inter-diffusion, appear to play vital roles in the electronic and transport properties of the metallic electron liquid at the oxide hetero-interfaces. Antisite disorder is also understood to be a key player in this context. In order to delineate the roles of two of these key factors, we have investigated the effect of oxygen vacancy on the antisite disorder, at a number of interfaces separating two La-based transition metal oxides, using density functional theory. Oxygen vacancy is found to suppress the antisite disorder in some heterostructures and thereby stabilizes the ordered structure, while in some others it tends to drive the disorder. Thus by controlling the oxygen partial pressure during the growth, it is possible to engineer the antisite disorder in many oxide heretostructures.

}
\end{abstract}

\maketitle

\section{Introduction}

Heterostructures of transition metal oxides (TMO) are particularly interesting as emergent, unusual states of matter are found at the interface\cite{antisite23} because two TMOs with different properties meet there. However, to grow atomically sharp interfaces is quite challenging and the physical properties of these systems get modified when defects are present at the interface \cite{antisite4,antisite5,antisite6,antisite7}. These defects are nearly impossible to eliminate and one has to understand their effects in detail. Any future application of these structures would crucially depend on our ability to predict and manipulate the nature and interplay of different kinds of disorders at the interface. 

TMOs have already been studied extensively during the last decade for their intriguing electronic properties, including metal-insulator transitions \cite{antisite1}, colossal magnetoresistance \cite{antisite2}, high T$_{c}$ superconductivity \cite{antisite3}. It is therefore not unexpected that their heterostructures are replete with rich physics. In many heterostructures, it is often energetically favorable for an atom to mutually interchange its position across the interface \cite{antisite8,antisite9,antisite10} with another. This antisite disorder is hard to identify in transmission electron microscopy, but appears to play a crucial role in controlling the near-interface properties. 

Again, depending on different growth and annealing conditions, oxygen vacancies (OV) can naturally occur at an oxide interface\cite{antisite21,antisite22}. An OV can trap electrons and may give rise to localized states \cite{antisite11}. It can also act as a source of charge carriers. The appearance of conducting interface between two band insulators LaAlO$_{3}$ and SrTiO$_{3}$ could be, according to some, down to the creation of OVs during the deposition of LaAlO$_{3}$ layers on the SrTiO$_{3}$ substrate \cite{antisite12,antisite13,antisite14}. The mobility of the 2D electron gas is found to increase with the lowering of the O-partial pressure and the conductivity gets suppressed as the O-partial pressure is increased. \cite{antisite15}\\

Anticipating the importance of atomically sharp interfaces in the context of Weyl semimetal and d-wave superconductivity\cite{antisite23,antisite24,antisite25}, Chen, \textit{et al}\cite{antisite16} have studied the prevalence of antisite disorder using \textit{ab-initio} calculations at the $RA$O$_{3}$/$RA'$O$_{3}$ type interfaces, where $R$ : La, Sr and $A$, $A'$ are the first row transition metals. They conclude that antisite disorder is energetically favorable in those heterostructures where there is a large strain generated due to lattice size mismatch of the participating oxide pairs. However, they do not consider O-vacancies at the interface. 
Since, oxygen vacancies can, and most probably do, occur naturally at oxide interfaces during the growth process, we present a study on the effect of O-site vacancy on the structural, magnetic and electronic properties of the La$A$O$_{3}$/La$A'$O$_{3}$ heterostructures. We also include antisite disorder and study the effect in a number of heterostructures.
We observe marked differences between results with and without the O-vacancy in several heterostructures and catalogue our results to delineate the differences. This could help in choosing the combinations and conditions for the growth processes of these heterostructures.  

The rest of the paper is organized as follows. In Section II we present the methodology adopted in our simulations. The crystal structure of the heterostructure is described in section III. We present our results in section IV. This is followed by discussion and conclusion in section V and VI respectively. 

\section{Methodology}

In the foregoing, we discuss results from density functional theory (DFT) calculations on a number of La$A$O$_{3}$/La$A'$O$_{3}$  heterostructures where $R$ : $A$, $A'$ are the first row transition metals. All calculations have been performed using first-principles density functional theory (DFT) as implemented in the plane wave based \textit{Vienna Ab-initio Simulation Package} (VASP)\cite{antisite17}. Generalized gradient approximation (GGA) of Perdew, Burke, and Ernzerhof (PBE) \cite{antisite18} is employed for the exchange-correlation part. The plane wave energy cutoff is set to 400 eV. Starting from the experimental data, we fully relax both the lattice parameters and the internal atomic co-ordinates until the Hellman-Feynman forces on each atom becomes smaller than 10meV/\r{A}. A $5 \times 5 \times 5$ Monkhorst-Pack is used for the full Brillouin zone. On-site correlation effects are taken into account in the GGA+U approximation 
\cite{antisite19} with values from literature \cite{antisite16}: U = 9.0 eV, J = 0.0 eV for the La-4f states. We take U = 5.0 eV for all transition metal 3d orbitals. However, for the 3d orbitals of early-transition metals (Ti, V, Cr) and late-transition metals (Mn, Fe, Ni, Co), J = 0.65eV and 1.00 eV are used respectively, as reported earlier \cite{antisite16}. We have also varied the value of U within reasonable range, but we get qualitatively similar results for different U values.\\ 

\section{Crystal structure}

La$A$O$_{3}$-type ($A$ : Ti, V, Cr, Mn, Fe, Ni, Co) compounds are known to crystalize in the orthorhombic space group Pbca (space group no. 62). Here each transition metal ion occupies the centre of the octahedron created by oxygen ions. In this study, a given oxide interface is constructed by considering a superlattice geometry with periodic repeatation of alternating layers of LaAO3 and LaA’O3 as shown in Fig. 1.

\begin{figure}[htp]
$\begin{array}{cc}
\includegraphics[scale=.32]{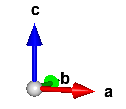}  &
\includegraphics[scale=.42]{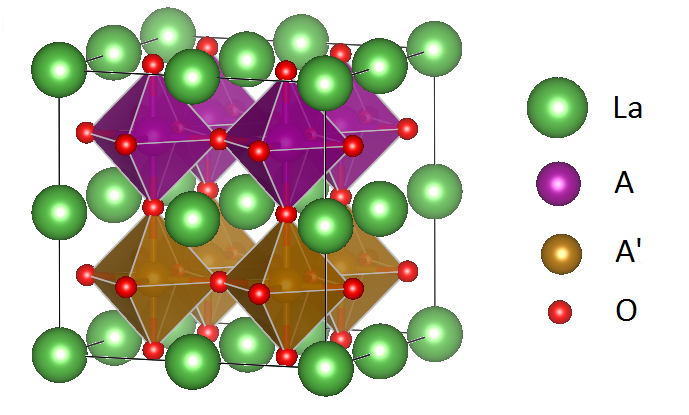} \\
\end{array}$
\caption{Supercell constructed using alternating layers of La$A$O$_{3}$ and La$A'$O$_{3}$.}
\end{figure}
\section{Results}
Considering all possible combinations of the first row transition metals, we obtain 21 different heterostructures of the form La$A$O$_{3}$/La$A'$O$_{3}$. 

In the following we present our studies on the ground state structural, magnetic and electronic properties of these heterostructures in the presence and absence of O-site vacancy. Since, TMOs are magnetic in nature, we have considered four possible magnetic configurations: 1) Ferromagnetic where all the spins are aligned in the same direction, 2) A-type antiferromagnetic where spins are aligned ferromagnetically in the plane but two consecutive planes are arranged antiferromagnetically, 3) C-type antiferromagnetic when spins are coupled antiferromagnetically in-plane and ferromagnetically out of plane and 4) checkerboard antiferromagnetic (G-type). We have shown the different magnetic configurations in Fig. 2. The ground state crystal structure and magnetic properties of a given heterostructure is obtained by comparing the total energies of the heterostructure with and without defect in all four possible magnetic configurations in all 21 heterostructures considered.
\begin{figure}[htp]
\centering
\includegraphics[scale=.33]{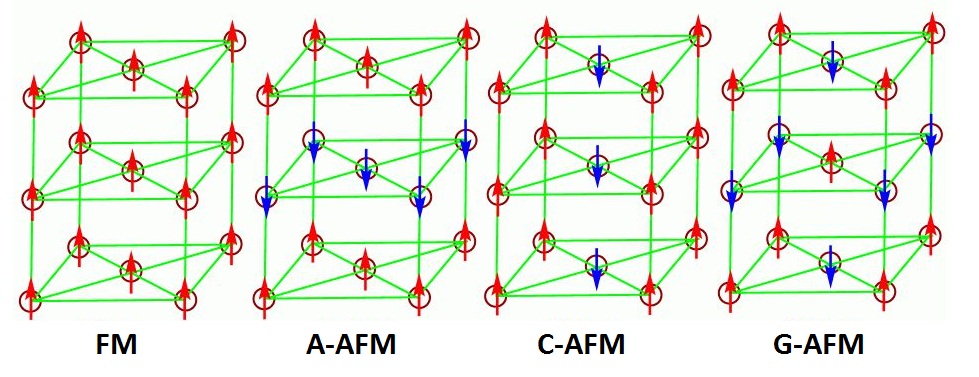}  
\caption{Different types of spin orderings}
\end{figure}

We further chcked the antisite defect and its stability in the presence of O vacancies. The antisite defect formation energy $\Delta E_{antisite}$  is defined as the energy difference between the lowest energy configuration in the presence of antisite defect ($E_{antisite}$) and the lowest energy ordered (no antisite defect) configuration($E_{order}$). This is done in all 21 heterostructures for 4 different magnetic configurations of Fig. 2. 

\begin{equation}
\Delta E_{antisite} = E_{antisite} - E_{order}
\end{equation}

In order to check the effect of lattice size, we have performed some preliminary calculations on a specific heterostructure, LaFeO$_{3}$/LaMnO$_{3}$, with two different in-plane dimension of the simulation cell, namely, $\sqrt2 \times \sqrt2$ (the same as Chen \textit{et al.}\cite{antisite16}) and a bigger one, $2 \times 2$ . It is to be noted that in a heterostructure having an in-plane dimension of $2 \times 2$, each layer contains four $A$O$_{6}$ octahedra as shown in Fig. 3. 
\begin{figure}[htp]
$\begin{array}{cc}
\includegraphics[scale=.30]{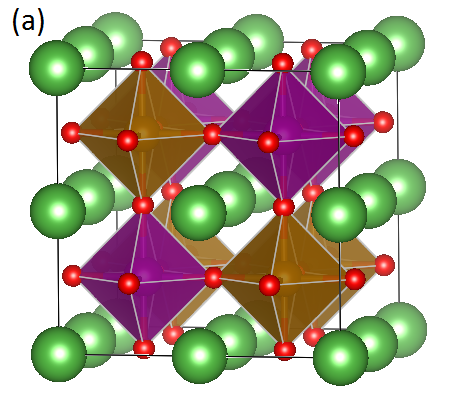} &
\includegraphics[scale=.28]{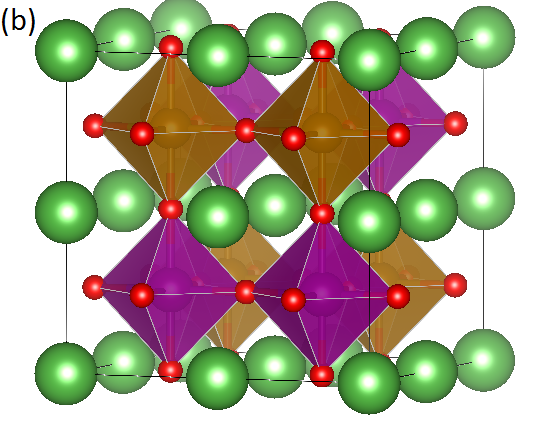} \\
\end{array}$
\caption{(a) Heterostructures with (a) 25\% and (b) 50\% antisite defects.}
\end{figure}
Therefore, flipping of one pair of transition metal ion implies 25\% defect concentration, whereas flipping of two TM ions implies 50\% defect concentration. Whereas, in a heterostructure constructed out of a simulation cell having a $\sqrt2 \times \sqrt2$ in-plane dimension, there are only two $A$O$_{6}$ octahedra in each layer of the heterostructure (as shown in Fig.3) and only 50\% antisite defect concentration can be studied. 
\begin{figure}[htp]
\centering
$\begin{array}{cc}
\includegraphics[scale=.20]{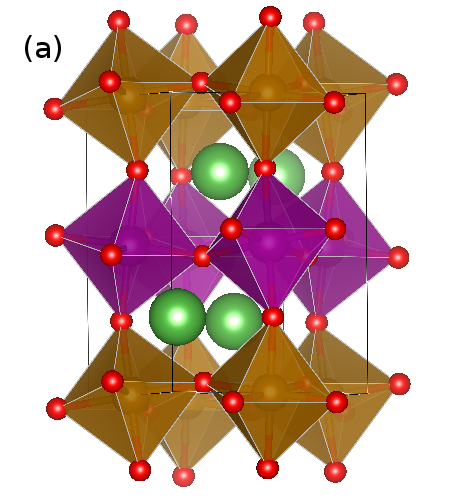}  &
\includegraphics[scale=.20]{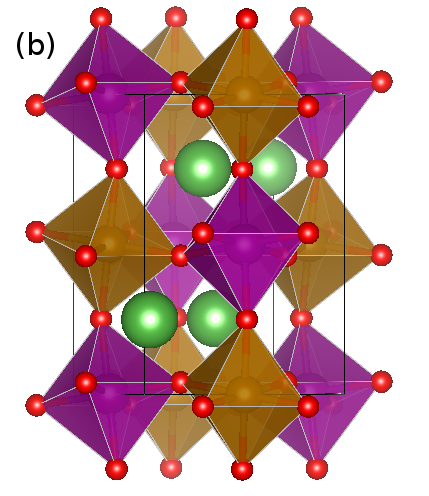} \\
\end{array}$
\caption{$\sqrt2 \times \sqrt2 \times 2$ cell (a) without any defect, (b) with 50\% antisite defect. Here, FeO$_{6}$ and MnO$_{6}$ octahedra are denoted by purple and yellow octahedra.}
\end{figure}

We have compared the ground state properties of 50\% antisite defect in simulation cells of (i) $\sqrt2 \times \sqrt2$  heterostructure and (ii) $2 \times 2$ heterostructure. The ordered structure is favorable when we use the $\sqrt2 \times \sqrt2$ type-supercell, similar to the findings of Chen \textit{et al.}\cite{antisite16} However, antisite disorder is seen to be energetically favorable when we use a $2 \times 2$ type supercell of the LaFeO$_{3}$/LaMnO$_{3}$ heterostructure with the same concentration of antisite defect. Calculated values of $\Delta E_{antisite}$ are shown in Table I. This shows that the cell size plays important role on the simulation results. We therefore use the larger $2 \times 2 \times 2$ supercell for all the calculations, which allows for greater degree of freedom when we perform relaxation of the lattice. 
\begin{table}[h]
\centering
\caption{Defect formation energies for different cell sizes}
\label{tab:Table2}
\begin{tabular}{|c|c|c|c|}
\hline
\hline
    \textbf{$A$} &  \textbf{$A'$} & \textbf{$\Delta E_{antisite}$ (eV)}  & \textbf{$\Delta E_{antisite}$ (eV)}\\
    \hline
                 &                & $\sqrt2 \times \sqrt2 \times 2$      & $2 \times 2 \times 2$\\
    \hline
    Fe & Mn & 0.17014 & -0.29272\\
    \hline
    \hline

  \end{tabular}
\end{table}

\subsection{Energetics}
\subsubsection{Without oxygen vacancy}

In this section we present our calculations on the antisite defect formation energy in the 21 different heterostructures of the form La$A$O$_{3}$/La$A'$O$_{3}$. As already stated, we have considered a $2 \times 2 \times 2$ supercell containing 40 atoms and 4 different magnetic structures. The defect concentration is fixed at 25\%. The value of $\Delta E_{antisite}$ is presented in Table II. 
\begin{table}[h]
\caption{Energetics in the absence of oxygen vacency. Here, $A$,$A'$ are the transition metal ions.}
\label{tab:Table1}
\begin{tabular}{|c|c|c||c|c|c|}
\hline
\hline
    \textbf{$A$} &  \textbf{$A'$} & \textbf{$\Delta E_{antisite}$ (eV)} & \textbf{$A$} &  \textbf{$A'$} & \textbf{$\Delta E_{antisite}$ (eV)}\\
    \hline
    V & Ti & -0.07292 & Co & V & 0.11117 \\
    Cr & Ti & 0.72168 & Co & Cr & 0.47923  \\
    Cr & V & -0.42961 & Co & Mn & 0.33165 \\
    Mn & Ti & -0.07265 & Co & Fe & -0.07086  \\
    Mn & V & -0.31137 & Ni & Ti & 0.21778 \\
    Mn & Cr & 1.03069 & Ni & V & 0.06112 \\
    Fe & Ti & 1.01954 & Ni & Cr & 0.84372  \\
    Fe & V & -0.94853 &  Ni & Mn & -0.63623  \\
    Fe & Cr & 0.15660 & Ni & Fe & 0.33788  \\
    Fe & Mn & -0.1702 &  Ni & Co & 1.83004  \\
    Co & Ti & 0.7604 & & &\\
    \hline
    \hline

  \end{tabular}
\end{table} 
\subsubsection{With oxygen vacancy}
In order to observe the effect of oxygen vacancy (OV) on the antisite disorder, we introduce one vacancy at the interface which corresponds to $\sim$4.2\% vacancy concentration, keeping the concentration of the defect (25\%) and the cell size ($2 \times 2 \times 2$) fixed. In this case, $\Delta E_{antisite}$ is calculated as
\begin{equation}
\Delta E_{antisite} = E^{\text{OV}}_{antisite} - E^{\text{OV}}_{order}
\end{equation}
where, $E^{\text{OV}}_{antisite}$ is the energy of the system with both antisite defect and OV and $E^{\text{OV}}_{order}$ is the energy of the ordered system which has a single vacancy at the interface. Fig.5 schematically shows the creation of an OV at the interface of ordered structures and heterostructures containing antisite defect respectively.
\begin{figure}[htp]
$\begin{array}{cc}
\includegraphics[scale=.28]{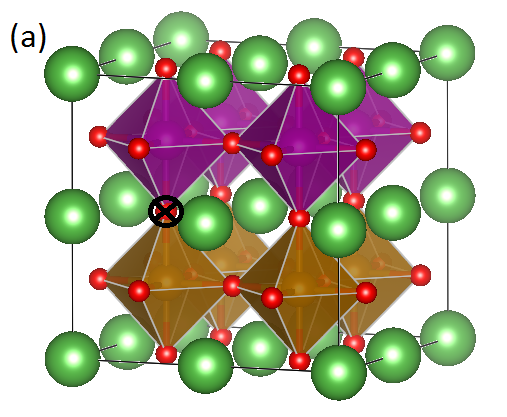} &
\includegraphics[scale=.29]{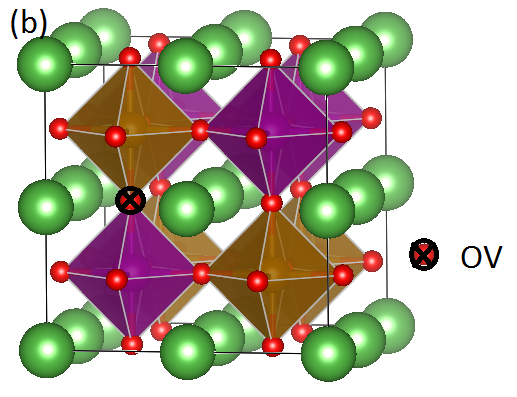} \\
\end{array}$
\caption{(a) Ordered structure with a single vacancy. (b) Supercell containing both antisite defect and OV.}
\end{figure}

Since all the interfacial oxygen sites are equivalent in the ordered structure, there is only one way to put the OV, anywhere in the interface. However, for the disordered structure, there are four possible ways in which the OV can be introduced at the interface for a given position of the antisite defect. In our work, we keep the positon of the defect fixed and create an OV at the four possible sites as shown in Fig. 6. It is to be noted that for each of these defect structures (as shown in Fig. 6), we have considered four possible magnetic configurations as elaborated in Fig. 2. While calculating $\Delta E_{antisite}$, $E^{\text{OV}}_{antisite}$ corresponds to the lowest energy defect structure in its most prefered magnetic configuration. Calculated defect formation energies in presence of OV are shown in Table III.\\
\begin{figure}[htp]
\centering
$\begin{array}{cc}
\includegraphics[scale=.30]{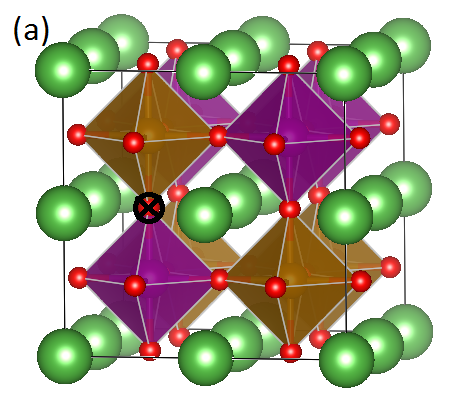} & 
\includegraphics[scale=.30]{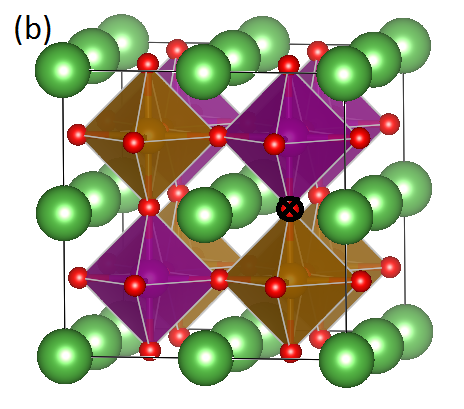}  \\
\includegraphics[scale=.30]{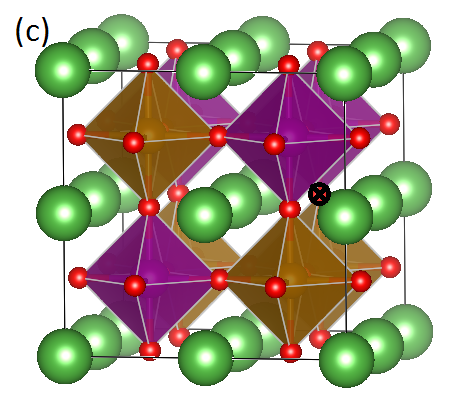} & 
\includegraphics[scale=.30]{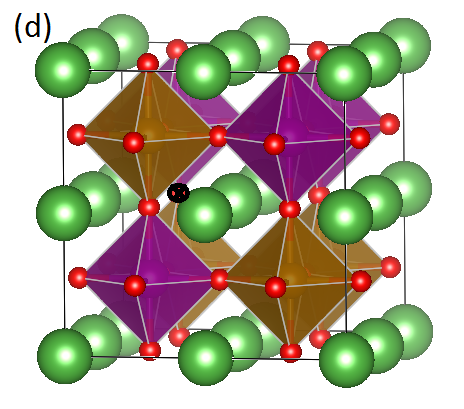}\\
\end{array}$
\caption{Disordered structure with an OV at the four different positions : (a) Position-1 (b) Position-2 (c) Position-3 (d) Position-4.}
\end{figure}

\begin{table}[h]
\centering
\caption{Energetics in presence of oxygen vacency. Position of the OV corresponds to Fig. 4(a), 4(b), 4(c) and 4(d).}
\label{tab:Table3}
\begin{tabular}{|c|c|c|c||c|c|c|c|}
\hline
\hline
    \textbf{$A$} &  \textbf{$A'$} & \textbf{$\Delta E_{antisite}$} & \textbf{Position} &  \textbf{$A$} &  \textbf{$A'$} & \textbf{$\Delta E_{antisite}$} & \textbf{Position}\\
    \hline
    V & Ti & 1.70884 &  - & Co & V & -1.40351  & 4\\
    Cr & Ti & -1.34554  & 3 & Co & Cr & -1.50503 & 4\\ 
    Cr & V & -0.54566 & 3 & Co & Mn & -0.85893  & 2\\
    Mn & Ti & -1.32485  & 3 & Co & Fe & -0.42689 & 2\\
    Mn & V & -0.24608  & 1 & Ni & Ti & 0.19104  & -\\
    Mn & Cr & -0.02194  & 2 & Ni & V & -0.09975 & 4\\
    Fe & Ti & 0.44403  & - & Ni & Cr & 0.23725  & - \\
    Fe & V & -1.53395  & 3 & Ni & Mn & -0.52059 & 4\\
    Fe & Cr & -0.32022  & 4 & Ni & Fe & 0.21421  & -\\
    Fe & Mn & 0.05659  & - & Ni & Co & -0.21706 & 3\\
    Co & Ti & -3.21332 & 4 & & & &\\
    \hline
    \hline
\end{tabular}
\end{table} 
\begin{figure*}
\centering
\onecolumngrid
\includegraphics[scale = 0.5]{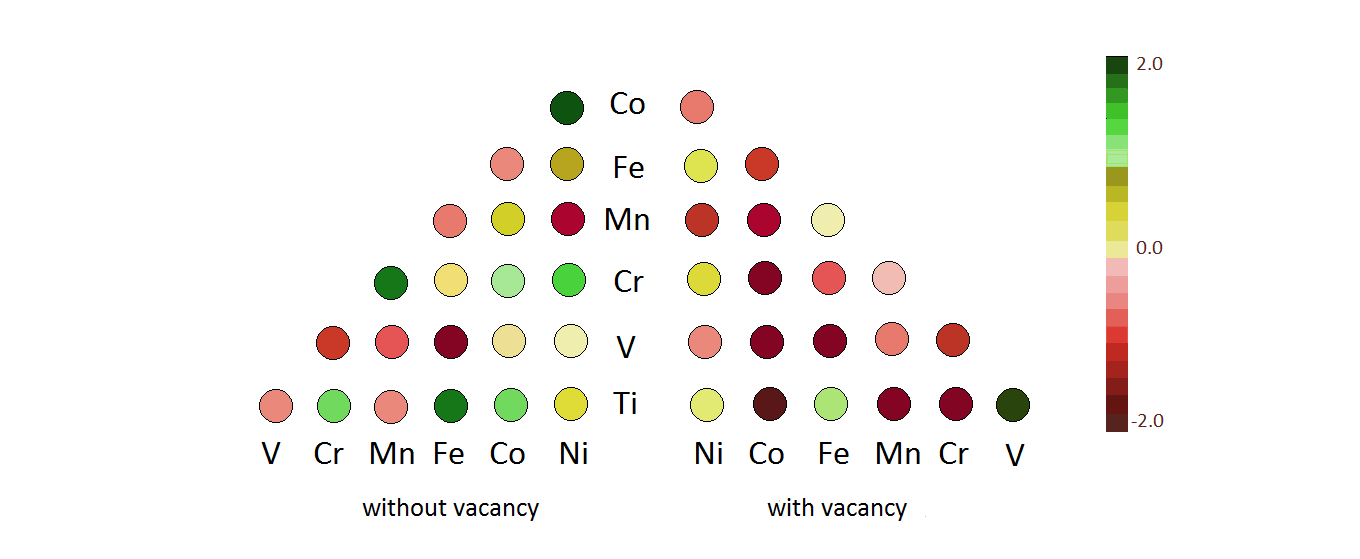}
\caption{Antisite defect formation energy (in eV) both in presence and in absence of oxygen vacancy. Positive values indicate that ordered structure is favored and vice-versa.}
\twocolumngrid
\end{figure*}

In Fig. 7, we summarise the results schematically. The color shading of the circles mean the following: red denotes  systems where the antisite disorder is favorable; starting from yellow onwards till green indicate that $\Delta E_{antisite}$ is positive (the ordered state has lower energy) i.e. the corresponding heterostructures are stable against disorder.

\subsection{Magnetic structure}

A comparison of the ground state electronic and magnetic structures of the oxide interfaces with and without vacancy is presented in Table IV. We find that the magnetic structures of the systems change when the OV is introduced. Creation of O-site vacancy causes a change in the bond angles, which in turn changes the exchange interactions between different transition metal atoms and as a result, the ground state spin orientations change.
\begin{table}
\caption{Ground state magnetic and electronic structures. Here, A-AFM, C-AFM, G-AFM and FM denote A type, C-type, G-type antiferromagnetic ordering and ferromagnetic ordering respectively. The metallic and insulating states are denoted by M and I.}
\label{tab:Table1}
\begin{tabular}{|c|c|c|c|}
\hline
\hline
    \textbf{$A$} &  \textbf{$A'$} & \textbf{without OV} & \textbf{with OV}\\
    \hline
    V & Ti & C-AFM(I) & C-AFM(I)\\
    Cr & Ti & G-AFM(M)& FM(M) \\
    Cr & V & C-AFM & A-AFM(I) \\
    Mn & Ti & FM(M) & G-AFM(I) \\
    Mn & V &  FM(M) &  G-AFM(I) \\
    Mn & Cr & A-AFM(M) & G-AFM(M)\\
    Fe & Ti & A-AFM(M) & A-AFM(M) \\
    Fe & V & G-AFM(M) & C-AFM(M)\\
    Fe & Cr & C-AFM(I) & G-AFM(I) \\
    Fe & Mn & C-AFM(I) & G-AFM(I) \\
    Co & Ti & G-AFM(I) & FM(I)\\
    Co & V & C-AFM & A-AFM(I)\\
    Co & Cr & G-AFM(I) & FM(I)  \\
    Co & Mn & FM(M) & FM(M)\\
    Co & Fe & G-AFM(I) & G-AFM(I) \\
    Ni & Ti & C-AFM(I) & C-AFM(M)  \\
    Ni & V & C-AFM(I) & G-AFM(I)\\
    Ni & Cr & C-AFM(M) & FM(M)  \\
    Ni & Mn & FM(M) & FM(M) \\
    Ni & Fe & G-AFM(M) & G-AFM(I) \\
    Ni & Co & A-AFM(M) & FM(M)\\
    \hline
    \hline
 \end{tabular}
\end{table} 
 
\subsection{Electronic structure}
Inclusion of oxygen vacancy at the interface not only changes the magnetic structures, but the electronic properties are also changed as shown in Table IV. In some systems, (e.g.LaNiO$_{3}$/LaTiO$_{3}$) the vacancy acts as a source of charge carriers and donates the extra two electrons making the insulating interface metallic. On the other hand, some metallic systems (e.g. LaMnO$_{3}$/LaTiO$_{3}$, LaMnO$_{3}$/LaVO$_{3}$) become insulating and the OV gives rise to localized states. In Table V, we list the systems where metal to insulator (MIT) and insulator to metal transition (IMT) takes place after the introduction of the vacancy. 
\begin{table}[h]
\centering
\caption{Heterostructures in which the OV changes the electronic structure.}
\label{tab:Table5}
\begin{tabular}{|c|c|c|}
\hline
\hline
    \textbf{Metal to insulator } &  \textbf{Insulator to metal} \\
    \hline
    \hline
    LaMnO$_{3}$/LaTiO$_{3}$ & \\
    LaMnO$_{3}$/LaVO$_{3}$ & LaNiO$_{3}$/LaTiO$_{3}$\\
    LaNiO$_{3}$/LaFeO$_{3}$ &\\
    \hline
    \hline
\end{tabular}
\end{table} 
\begin{figure*}
\centering
\onecolumngrid
$\begin{array}{cccc}
\includegraphics[scale=.31]{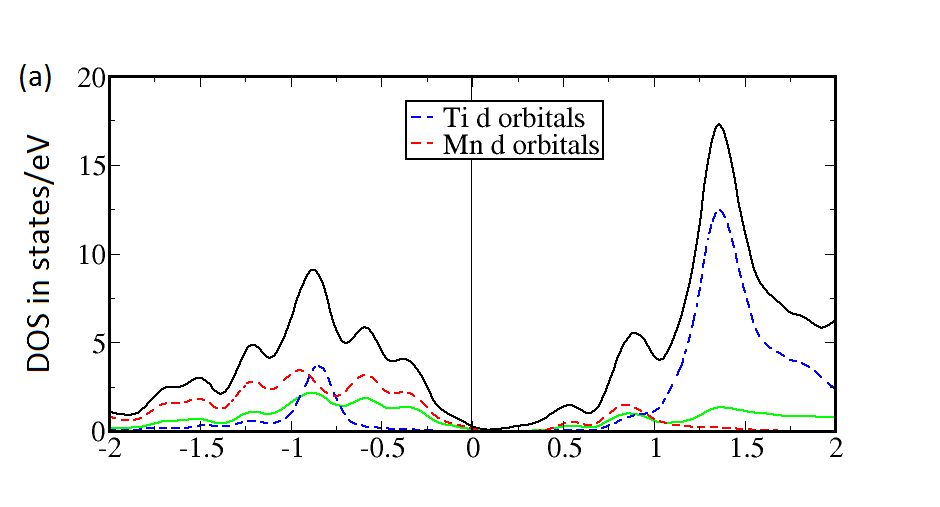}  &
\includegraphics[scale=.31]{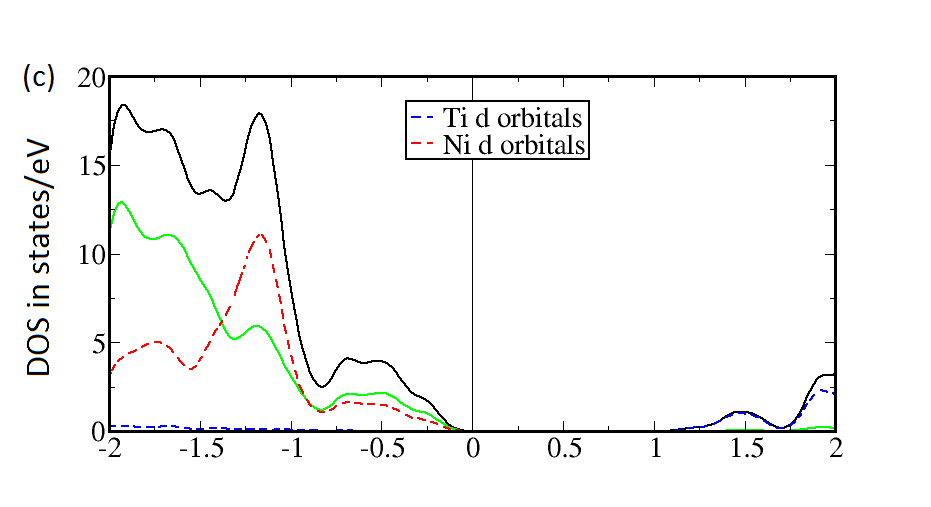} \\
\includegraphics[scale=.31]{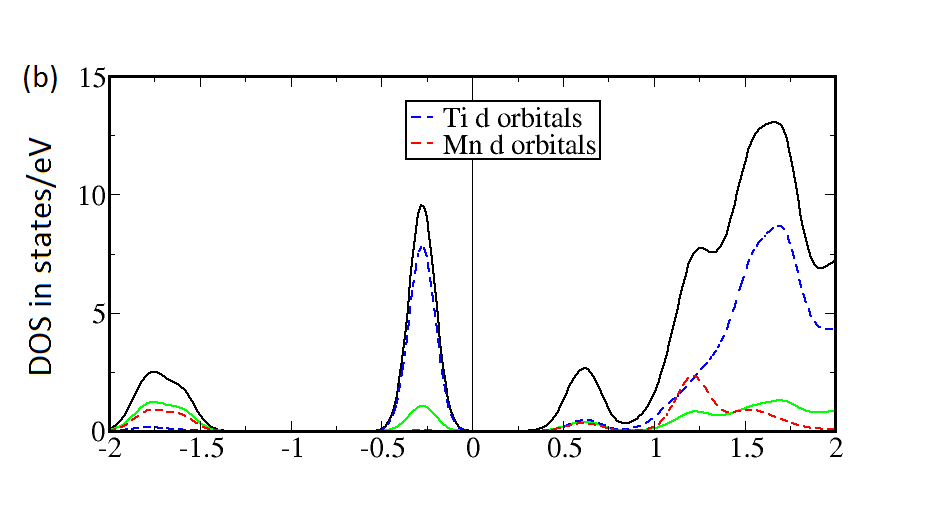} &
\includegraphics[scale=.31]{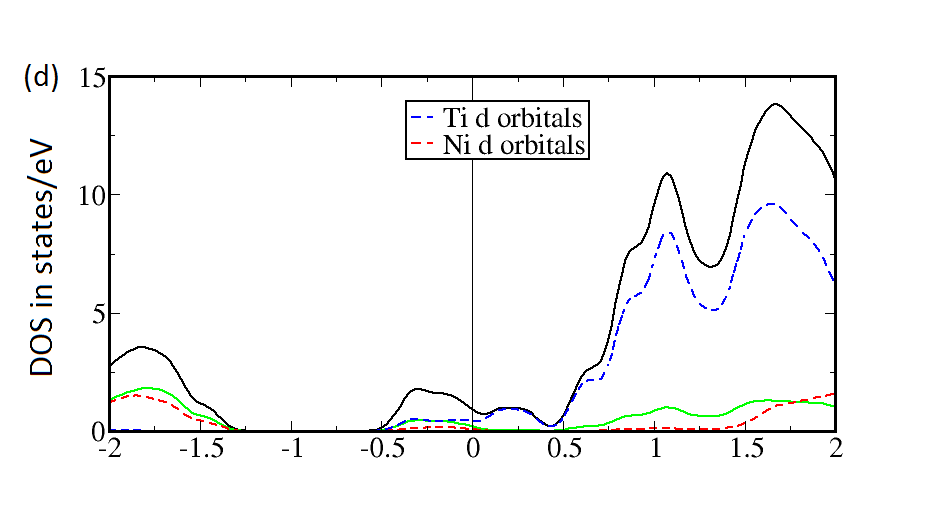} \\
\end{array}$
\caption{Density of states (DOS) (plotted as a function of energy in eV). Here, black bold lines represent total DOS and green bold lines represent O-p partial DOS. (a) LaMnO$_{3}$/LaTiO$_{3}$ heterostructure without O-vacancy and (b) with O-vacancy. (c) LaNiO$_{3}$/LaTiO$_{3}$ heterostructure without O-vacancy and (d) with O-vacancy.}
\twocolumngrid
\end{figure*}
Fig. 8 shows the density of states of two systems where the MIT and Insulator-metal transition (IMT) take place. In the case of LaMnO$_{3}$/LaTiO$_{3}$, there is a tiny density of states near the Fermi level when the system does not contain any vacancy. However, the introduction of O-vacancy produces a well-defined energy gap $\sim$ 0.26 eV. On the other hand, LaNiO$_{3}$/LaTiO$_{3}$ shows insulating behavior without OV with a band gap of $\sim$ 1.1 eV which, however, closes on the introduction of OV at the interface. 
\begin{figure*}
\centering
\onecolumngrid
$\begin{array}{ccc}
\includegraphics[scale=.32]{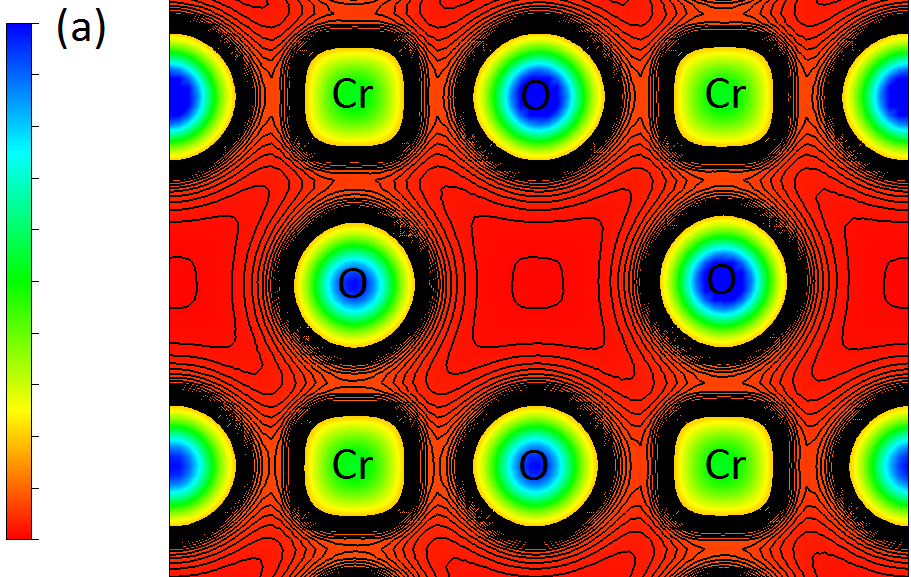} & 
\includegraphics[scale=.32]{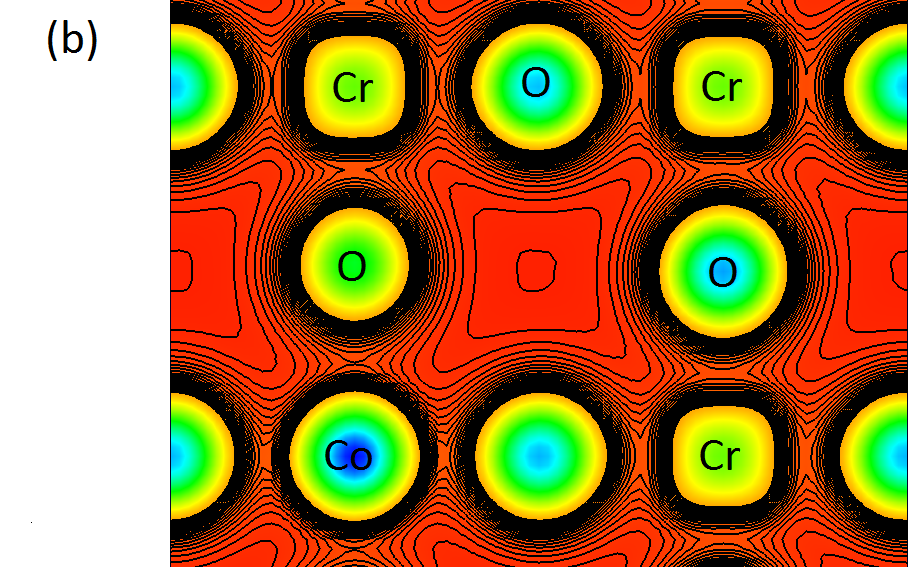}  \\
\includegraphics[scale=.32]{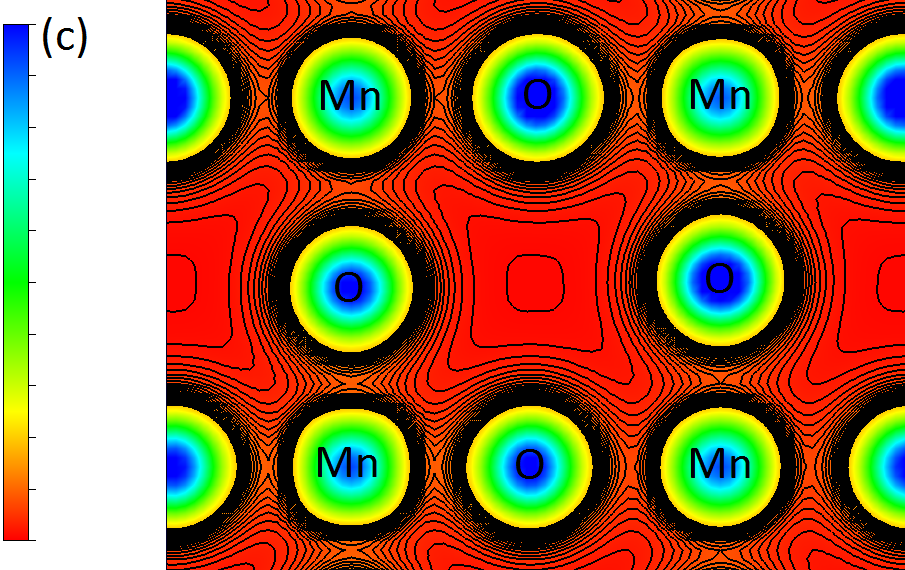} & 
\includegraphics[scale=.32]{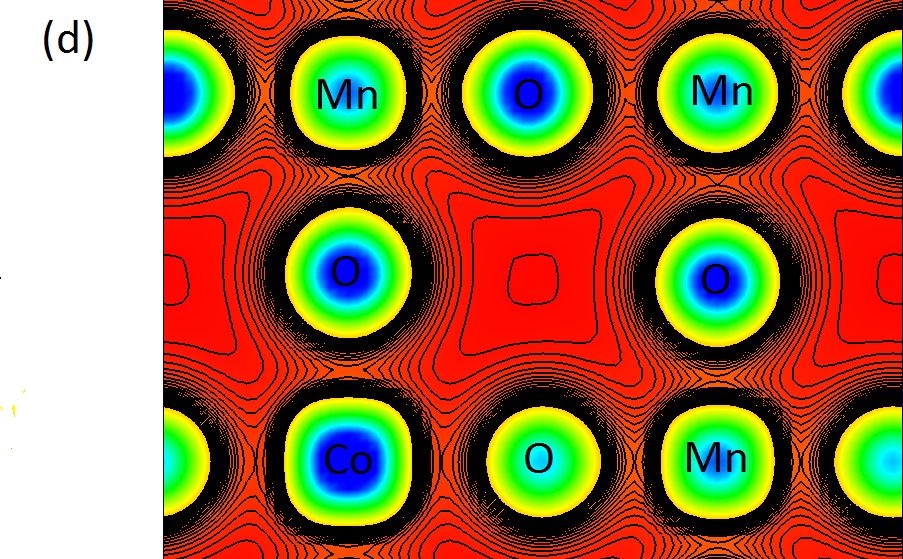}\\
\includegraphics[scale=.32]{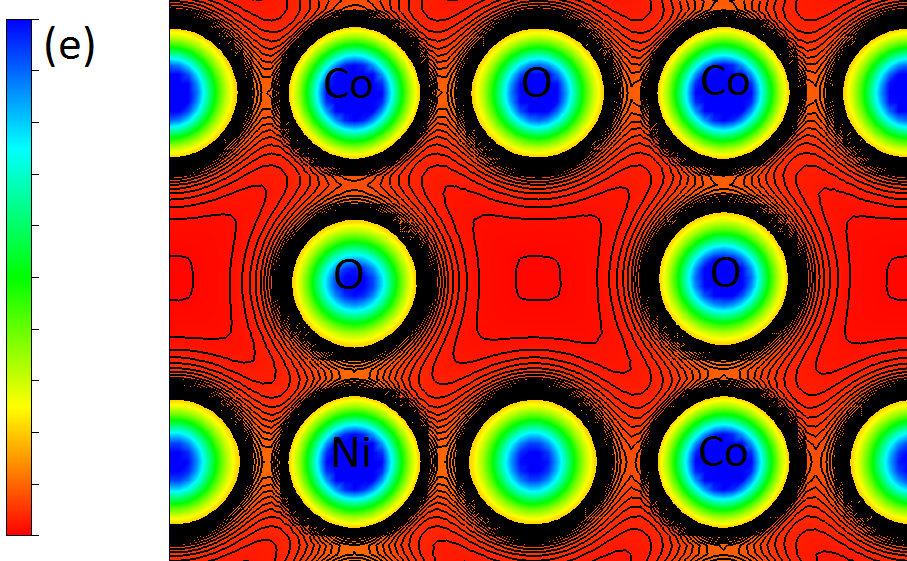} & 
\includegraphics[scale=.32]{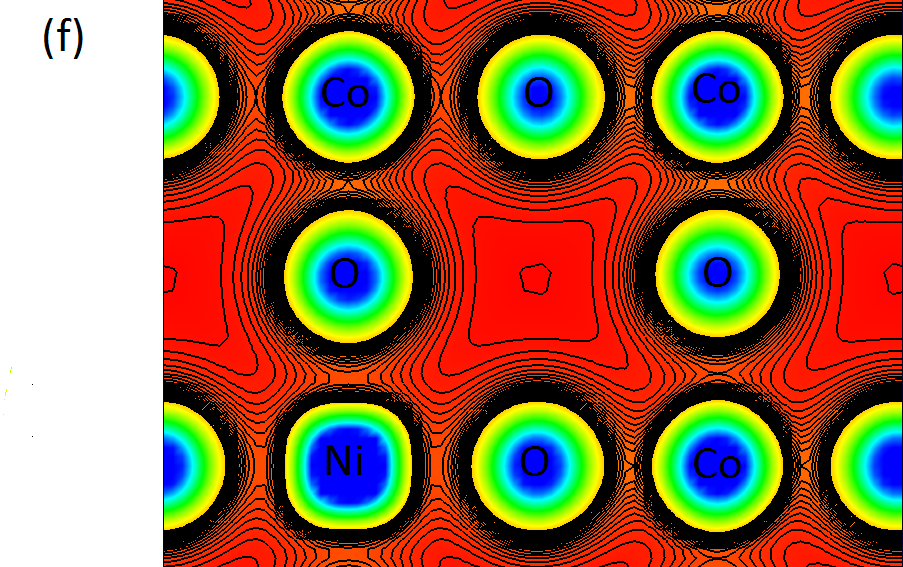}\\
\end{array}$
\caption{Charge density plots projected onto the $A'$O$_{6}$ plane in presence of oxygen vacancy : (a),(c),(e) are for the ordered structures and (b),(d),(f) are the charge densities for the disordered structures.}
\twocolumngrid
\end{figure*}

\section{Discussion}
In Table VI we enlist the systems in which the presence of oxygen vacancy at the interface affects the ordering of the transition metal cations. We find that there is a class of systems in which the antisite disorder is energetically favored in the absence of OV. However, inclusion of OV at the interface stabilizes the ordered structures. Similarly, there is a class of heterostructures for which the OV drives the antisite disorder, which were stable against the disorder in the absence of the OV. Therefore, we see that the introduction of OV can influence antisite disorder across heterostructure interface. 

In general, in a given La$A$O$_{3}$/La$A'$O$_{3}$ type heterostructure, there are two driving factors, namely lattice strain and covalency, which should drive order/disorder across the interface. Lattice strain comes into play because of the mismatch in the cationic radii of the two transition metals $A$ and $A'$. Whereas, very strong hybridization between the p-orbitals of oxygen ion and the d-orbitals of transition metal ion (as can be seen from the density of states plot in Fig. 8) indicates that these oxides are highly covalent. A given system will hence try to minimize strain and maximize covalency in order to attain stability. In this study, we have considered the volume of $A$O$_{6}$ octahedra as an approximate measure of the cationic radii of $A$. Whereas, the difference in the octahedral volume of $A$O$_{6}$ and $A'$O$_{6}$ octahedra is indicative of the lattice strain generated due to the formation of the heterostructure. In Table VII, we present the octahedral volume difference (i.e, the difference in the average $A$O$_{6}$ and $A'$O$_{6}$ octaheral volume) of the energetically most stable structure in the presence and absence of vacancy for those systems in which oxygen vacancy influences order/disorder across the interface. We find that except for three cases, namely LaCoO$_{3}$/LaCrO$_{3}$, LaCoO$_{3}$/LaMnO$_{3}$ and LaCoO$_{3}$/LaNiO$_{3}$, the ground state crystal structure of the heterostructure is the one which has the minimum lattice strain, measured in terms of the difference in the volume of the $A$O$_{6}$ and $A'$O$_{6}$ octahedra. 

For example, in LaCoO$_{3}$/LaCrO$_{6}$, the difference between the average octahedral volume in the Co-plane and the average octahedral volume in the Cr-plane is 0.3782. In the presence of antisite disorder, the  same becomes 0.5739. That is, in the absence of vacancy, the octahedral volume difference between the octahedra belonging to the two planes sharing the interface is minimum in the ordered structure. Therefore the system is expected to remain in an ordered state in the absence of OV, in order to minimize strain in the lattice. Total energy calculations also suggest the same. 

However, in the presence of vacancy we find that although the difference between the octahedral volume is less when the system is in ordered state (0.0412), as oppose to the disordered case (0.3163), total energy calculations suggest that the structure actually prefers to have antisite disorder. We find that with the introduction of OV, the difference of the volume difference between $A$O$_{6}$ and $A'$O$_{6}$ octahedra between the ordered and disordered state increases. This implies that the strain in the disordered structure relative to the ordered structure increases with introduction of OV. Analysis of the crystal structure of LaCoO$_{3}$/LaMnO$_{3}$ in the presence and the absence of vacancy reveals similar trends. 

The case of LaNiO$_{3}$/LaCoO$_{3}$ is somewhat different. Here, although the strain is less for the ordered structure in the case where there is OV, the strain in the disordered structure relative to the ordered structure seems to decrease with the introduction of OV. In other words, the strains in the ordered and disordered cases become comparatively similar in the presence of OV, as opposed to when there was no vacancy. The strain is again measured from the average octahedral volume difference between the two planes.  Therefore, we find that in all these cases, although the octahedral volume difference is minimum in the ordered case, total energy calculations suggest that antisite disorder is energetically favorable, thus giving an impression that the system prefers to remain in a strained state, which is indeed counter-intutive. 

Moreover, the strain in such oxide heterostructure lattice can also relax either by distortion or by rotation of the octahedra. Creation of OV is expected to make the octahedra distorted. We measure this distortion is terms of standard deviation in the average volume. We find that in the ordered state, for all these cases, the standard deviations in the average volume of the $A$O$_{6}$ and $A'$O$_{6}$ octahedra are similar. However, in the presence of antisite disorder, in the case of LaCoO$_{3}$/LaCrO$_{3}$ and LaCoO$_{3}$/LaMnO$_{3}$, not only does the standard deviation increase, the value of it for each plane becomes very different from the other. For example, in LaCoO$_{3}$/LaCrO$_{3}$, the standard deviation in the average volume of the octahedra in the Co-plane and in the Cr-plane is 0.27A and 0.28A respectively. However, with the introduction of OV, the standard deviation of the CoO$_{6}$ octahedra in the Co-plane and CrO$_{6}$ octahedra in the Mn-plane become 0.54 and 0.90 respectively. Similarly, in the case of LaCoO$_{3}$/LaMnO$_{3}$, the standard deviation in the CoO$_{6}$ octahedra in the Co-plane and the MnO$_{6}$ octahdedra in the Mn-plane of the ordered structure is 0.30 and 0.23 respectively, which become 0.73 and 0.34 with $OV$. In the case of LaNiO$_{3}$/LaCoO$_{3}$, the standard deviation in the average volume of NiO$_{6}$ octahedra and CoO$_{6}$ octahedra in the ordered structure is 0.23 and 0.24 respectively, which changes to 0.12 and 0.30 respectively with OV. Therefore, we find that in the OV containing LaCoO$_{3}$/LaCrO$_{3}$ and LaCoO$_{3}$/LaMnO$_{3}$ systems, presence of antisite disorder enhances the distortion of the lattice, which is turn can reduce the strain. 

Lastly, we also calculate the charge density of the minimum energy ordered configuration and minimum energy disordered configuration for  LaCoO$_{3}$/LaCrO$_{3}$ , LaCoO$_{3}$/LaMnO$_{3}$ and LaCoO$_{3}$/LaNiO$_{3}$. As can be seen from Fig. 9, covalency is enhanced when one of the atoms is exchanged across the interface in presence of oxygen vacancy. Therefore covalency works hand-in-hand with distortions in the octahedra to make the antisite diordered state stable in these systems.
\begin{table}[h]
\centering
\caption{Heterostructures where the OV suppresses the antisite disorder}
\label{tab:Table4}
\begin{tabular}{|c|c|}
\hline
\hline
    \textbf{Disordered to ordered} & \textbf{Ordered to disordered} \\
    \hline
     			    & LaCrO$_{3}$/LaTiO$_{3}$\\
                            & LaMnO$_{3}$/LaCrO$_{3}$\\
    LaVO$_{3}$/LaTiO$_{3}$  & LaFeO$_{3}$/LaCrO$_{3}$\\
    LaFeO$_{3}$/LaMnO$_{3}$ & LaCoO$_{3}$/LaTiO$_{3}$\\
                            & LaCoO$_{3}$/LaVO$_{3}$\\
                            & LaCoO$_{3}$/LaCrO$_{3}$\\
                            & LaCoO$_{3}$/LaMnO$_{3}$\\ 
                            & LaNiO$_{3}$/LaVO$_{3}$\\
                            & LaNiO$_{3}$/LaCoO$_{3}$\\

    \hline
    \hline
\end{tabular}
\end{table}

\begin{table}
    \begin{tabular}{|c|c|c|c|c|c|}
    \hline
    \hline
      &  & \multicolumn{4}{c|}{Octahedral volume differences} \\ \cline{3-6} 
      & & \multicolumn{2}{c|}{Without $OV$}& \multicolumn{2}{c|}{With $OV$} \\ \hline
    $A$ & $A'$ & Ordered & Disordered & Ordered & Disordered \\ \hline
     V & Ti & 1.4523 & 1.1059 & 0.5120 & 1.1524 \\ 
    Fe & Mn & 1.0127 & 0.1491 & 0.4768 & 0.5863 \\ 
    Cr & Ti & 0.2661 & 0.9182 & 0.4970 & 0.1266 \\
    Mn & Cr & 0.0202 & 0.0845 & 0.2203 & 0.2179 \\
    Fe & Cr & 0.0089 & 0.0143 & 0.0603 & 0.0873 \\
    Co & Ti & 0.6640 & 0.8990 & 0.7484 & 0.3475 \\
    Co & V & 1.0913 & 1.8232 & 0.9273 & 0.8558 \\
    Co & Cr & 0.3782 & 0.5739 & 0.0412 & 0.3163 \\
    Co & Mn & 0.6761 & 0.8169 & 0.4403 & 0.8091 \\
    Ni & V & 0.7948 & 1.0784 & 0.6491 & 0.3972 \\
    Ni & Co & 0.4619 & 1.0998 & 0.0948 & 0.4304 \\ \hline \hline
\end{tabular}
    \caption{Volume differences between $A$O$_{6}$ and $A'$O$_{6}$ octahedra in the ordered and disordered structures in presence and in absence of oxygen vacancy.}
 \end{table}

\section{Conclusion}
In conclusion, we have studied 21 La-based oxide heterostructures (La$A$O$_{3}$/La$A'$O$_{3}$) employing first-principles calculations. Antisite defect formation energy is calculated both in absence and in presence of oxygen vacancy. Our calculations reveal that oxygen vacancy plays an important role in controlling the antisite disorder at oxide interfaces. \textit{OV} prevents this disorder in a number of such systems. For few other systems, it favors the disordered structures. Therefore, the defects can be stabilized or destabilized depending on the oxygen partial pressure during the growth of these kind of systems, which motivates further experimental studies on these heretostructures. Thus the creation of oxygen site vacancy at the interface provides a new and unexplored opportunity to control antisite defect across heterostructure boundaries.   

\section{Acknowledgements}
We acknowledge research funding from CSIR (India) through grant number: 03(1373)/16/EMR-II. AT acknowledges D.D.Sarma for useful discussions.

\bibliography{vacancy}

\begin{thebibliography}{10}

\bibitem{antisite23}
J.~Chakhalian and J.~W. Freeland and A.~J. Millis and C.~Panagopoulos and J.~M.
  Rondinelli,
\newblock Reviews of Modern Physics {\bf 86}, 1189 (2014).

\bibitem{antisite4}
H.~Sato and C.~Bell and Y.~Hikita and H.~Hwang,
\newblock Applied Physics Letters {\bf 102}, 251602 (2013).

\bibitem{antisite5}
N.~Nakagawa and H.~Y. Hwang and D.~A. Muller,
\newblock arXiv preprint cond-mat/0510491  (2005).

\bibitem{antisite6}
J.~Jeong and N.~Aetukuri and T.~Graf and T.~D. Schladt and M.~G. Samant and
  S.~S. Parkin,
\newblock Science {\bf 339}, 1402 (2013).

\bibitem{antisite7}
N.~Mohanta and A.~Taraphder,
\newblock Journal of Physics: Condensed Matter {\bf 26}, 215703 (2014).

\bibitem{antisite1}
M.~Imada and A.~Fujimori and Y.~Tokura,
\newblock Reviews of modern physics {\bf 70}, 1039 (1998).

\bibitem{antisite2}
M.~B. Salamon and M.~Jaime,
\newblock Reviews of Modern Physics {\bf 73}, 583 (2001).

\bibitem{antisite3}
P.~A. Lee and N.~Nagaosa and X.-G. Wen,
\newblock Reviews of modern physics {\bf 78}, 17 (2006).

\bibitem{antisite8}
G.~Tuttle and H.~Kroemer and J.~H. English,
\newblock Journal of Applied Physics {\bf 67}, 3032 (1990).

\bibitem{antisite9}
P.~Lin-Chung and T.~Reinecke,
\newblock Journal of Vacuum Science and Technology {\bf 19}, 443 (1981).

\bibitem{antisite10}
R.~Allen and J.~Low,
\newblock Solid State Communications {\bf 45}, 379 (1983).

\bibitem{antisite21}
M.~Basletic and J.-L. Maurice and C.~Carretero and G.~Herranz and O.~Copie and
  M.~Bibes and E.~Jacquet and K.~Bouzehouane and S.~Fusil and A.~Barthelemy,
\newblock Nat Mater {\bf 7}, 621 (2008).

\bibitem{antisite22}
J.~Park {\em et~al.},
\newblock Physical review letters {\bf 110}, 017401 (2013).

\bibitem{antisite11}
H.~C. Nguyen and J.~B. Goodenough,
\newblock Physical Review B {\bf 52}, 8776 (1995).

\bibitem{antisite12}
A.~Ohtomo and H.~Hwang,
\newblock Nature {\bf 427}, 423 (2004).

\bibitem{antisite13}
W.~Siemons and G.~Koster and H.~Yamamoto and W.~A. Harrison and G.~Lucovsky and
  T.~H. Geballe and D.~H. Blank and M.~R. Beasley,
\newblock Physical review letters {\bf 98}, 196802 (2007).

\bibitem{antisite14}
A.~Kalabukhov and R.~Gunnarsson and J.~B{\"o}rjesson and E.~Olsson and
  T.~Claeson and D.~Winkler,
\newblock Physical Review B {\bf 75}, 121404 (2007).

\bibitem{antisite15}
G.~Herranz {\em et~al.},
\newblock Physical review letters {\bf 98}, 216803 (2007).

\bibitem{antisite24}
S.~Adam and E.~Hwang and V.~Galitski and S.~D. Sarma,
\newblock Proceedings of the National Academy of Sciences {\bf 104}, 18392
  (2007).

\bibitem{antisite25}
S.~D. Sarma and S.~Adam and E.~Hwang and E.~Rossi,
\newblock Reviews of Modern Physics {\bf 83}, 407 (2011).

\bibitem{antisite16}
H.~Chen and A.~Millis,
\newblock Physical Review B {\bf 93}, 104111 (2016).

\bibitem{antisite17}
G.~Kresse and J.~Furthm{\"u}ller,
\newblock Physical review B {\bf 54}, 11169 (1996).

\bibitem{antisite18}
J.~P. Perdew and K.~Burke and M.~Ernzerhof,
\newblock Physical review letters {\bf 77}, 3865 (1996).

\bibitem{antisite19}
A.~Liechtenstein and V.~Anisimov and J.~Zaanen,
\newblock Physical Review B {\bf 52}, R5467 (1995).

\end{thebibliography}

\end{document}